\documentclass[aps,prd,twocolumn,showpacs,preprintnumbers,amsmath,nofootinbib]{revtex4}

\usepackage[english]{babel}
\usepackage{amsmath,amssymb}
\usepackage{graphicx}
\usepackage[sort&compress]{natbib}
\usepackage{slashed}            
\allowdisplaybreaks

\bibpunct{[}{]}{,}{n}{}{,}

\def\lsim{\mathrel{\rlap{\lower4pt\hbox{\hskip1pt$\sim$}}
    \raise1pt\hbox{$<$}}}                
\def\gsim{\mathrel{\rlap{\lower4pt\hbox{\hskip1pt$\sim$}}
    \raise1pt\hbox{$>$}}}                

\def\be{\begin{equation}}
\def\ee{\end{equation}}
\def\bea{\begin{eqnarray}}
\def\eea{\end{eqnarray}}

\usepackage{xcolor}
\definecolor{red}{rgb}{1.0, 0, 0}

\begin{document}

\preprint{FERMILAB-PUB-13-024-T}

\title{Finding the Higgs boson in decays to $Z\gamma$ using the matrix element method at Next-to-Leading Order.}
\author{John M. Campbell}          \email[Email: ]{johnmc@fnal.gov}
\author{R. Keith Ellis}             \email[Email: ]{ellis@fnal.gov}
\author{Walter T. Giele}          \email[Email: ]{giele@fnal.gov}
\author{Ciaran Williams}            \email[Email: ]{ciaran@fnal.gov} \vspace{0.2cm}

\affiliation{Theoretical Physics Department \\
                  \mbox{Fermilab,  P.O.~Box 500, Batavia, IL 60510, USA} \vspace{0.2cm}\\
 }

\begin{abstract}
We illustrate how the Matrix Element Method at Next-to-Leading Order (MEM@NLO) can be used to discriminate between events arising  from
the production of a Higgs boson, which subsequently decays to a final state consisting of $\ell^+\ell^-\gamma$, and the background
production of the same final state. We illustrate how the method could be used in an experimental analysis by devising cuts
on the signal $(P_S)$ and background $(P_B)$ weights that are computed event-by-event in this approach.
We find that we can increase the
$S/\sqrt{B}$ ratio by around $50\%$ compared to an invariant mass fit on its own. Considering only statistical uncertainty, this is
equivalent to recording a factor of around two times more integrated luminosity.    
\end{abstract}
\date{\today}

\maketitle

\section{Introduction} 

The recent discovery of a new boson with properties consistent with
that of the Standard Model (SM) Higgs boson~\cite{Aad:2012gk,cms:2012gu},
has indicated that the discovery of the electroweak (EW) symmetry
breaking mechanism may be at hand. In order to confirm whether the new
boson is indeed the SM Higgs, it is crucial to measure both its properties
and branching ratios for the largest number of experimentally viable
decay channels. These analyses could result in tension with the SM Higgs prediction, 
for instance the boson may differ in parity from the SM Higgs or even be a mixture of CP odd and even states. 
An additional possibility is that the rate for one or more measured decay channels is different 
from the SM prediction. The most obvious mechanism for such a scenario is an enhancement/suppression in loop-induced decays that
are naturally sensitive to couplings to new virtual particles, for instance the decay to two photons 
($\gamma\gamma$). This would thus be evidence for Beyond the Standard Model (BSM) physics. 

Another loop-induced Higgs decay is the decay to a final state
containing a $Z$ boson and a photon ($H\rightarrow Z\gamma$)~\cite{Cahn:1978nz,Bergstrom:1985hp}. Since it is a 
loop-induced process the branching ratio is small and the decay
of the $Z$ boson to well-measured final state particles ($\mu^+\mu^-$
or $e^+e^-$) means that the decay $H\rightarrow \ell^+\ell^-\gamma$ is a very rare SM
process. However, this is not necessarily the case in extensions of the SM.
In addition, the ratio of $\gamma\gamma$ to
$Z\gamma$ branching ratios can be used to discriminate between certain
models of new physics~\cite{Chiang:2012qz,Chen:2013vi,Cao:2013ur}.

Observing the Higgs in the $Z\gamma$ final state is a difficult
feat. Firstly, the background production of $Z\gamma$ is several orders of
magnitude larger than the Higgs induced rate. The exact value of this
ratio depends on the cuts defining the background cross section. In the region of invariant 
mass near the Higgs mass ($m_{\ell\ell\gamma} \sim 125$~GeV), with
typical LHC cuts one would usually expect around 500-1000 background
events for each signal event.  Secondly,
the kinematics of the decay limit the final state photon to a challenging region of
phase space. At Leading-Order (LO) the maximum transverse momentum ($p_T$) of the
final-state photon is restricted, since the final-state invariant mass
is close to $m_H$ and includes a pair of charged leptons of
mass close to the $Z$ mass ($m_Z$). Since $m_Z$ is not too far from $m_H$, the remaining
energy to be imparted to the photon lies in a limited
range. Therefore the $p_T$ spectrum of the photon peaks around 30
GeV. 

Typically in such a soft region of phase space, QCD can provide large backgrounds to searches. 
This increases the difficulty in separating signal from
background when compared with $H\rightarrow\gamma\gamma$, for which the
photon $p_T$ from the signal is significantly harder
($p_T^{\gamma}\sim 60$~GeV). Indeed, once the full detector simulation
has been included, there are only small differences between the signal
and background shapes in the transverse variables~\cite{CMS:Zgam}. In terms of 
final state kinematics, the main discriminating variable is the angle between the direction of the 
photon and the beam ~\cite{Gainer:2011aa,Choi:2012yg}. A spin-0 scalar is isotropic in this variable, whilst 
the background matrix element prefers emission in the forward region.  
A recent CMS study with around 10 fb$^{-1}$ of 7 and 8 TeV data~\cite{CMS:Zgam}, set a
limit around 10 times the SM cross section. This result already disfavors scenarios in which the new boson 
is a pure pseudo-scalar since in some of these models the branching ratio to $Z\gamma$ can be enhanced by up to $170$ times
the SM prediction~\cite{Coleppa:2012eh}.

An experimental search for $H\rightarrow Z\gamma$ should thus utilize
as much theoretical information as possible in order to effectively
reduce the unwanted background $Z\gamma$ events. One such method is the Matrix
Element Method (MEM)~\cite{Kondo:1991dw,Dalitz:1991wa,Alwall:2010cq,Artoisenet:2010cn,Volobouev:2011vb}. This
method uses the matrix element associated with a given theoretical
hypothesis to assign a probabilistic weight to an experimental
event. Comparing weights obtained by varying the theoretical
hypothesis allows one to identify the most favorable
one. Originally the method was used in order to
perform a measurement of a known theoretical parameter, say for
instance the top mass~\cite{Abazov:2004cs,Abazov:2006bd,Abulencia:2006ry,Abulencia:2007br}. More recently, the method
has gained favor as an event by event
discriminant~\cite{Gainer:2011xz,Gainer:2011aa,Andersen:2012kn,Freitas:2012uk,CMS:2012br,Avery:2012um}. By
using the matrix element one naturally includes all of the
kinematic correlations present in the observed final state, and thus
gains a large amount of theoretical information. Until recently a
major drawback of the MEM was its restriction to LO matrix
elements. However in ref.~\cite{Campbell:2012cz} a new version of
the MEM was proposed that can be extended to
higher orders in perturbation theory\footnote{With the caveat that the
final state of interest should consist of only EW particles}.
Using the NLO method provides a
much greater degree of theoretical reliability and control
over the theoretical systematics.

Recently the MEM has been used in searches for, and studies of, the Higgs boson. The MEM at
LO has been applied to Higgs searches in the $ZZ$~\cite{Gainer:2011xz} and
$Z\gamma$~\cite{Gainer:2011aa} channels.
The MEM has also recently been applied to study the properties 
of the Higgs (decaying to two photons) via vector boson fusion~\cite{Andersen:2012kn}
and to investigate its role in unitarizing $WW$ scattering~\cite{Freitas:2012uk}.
The $Z\gamma$ search~\cite{Gainer:2011aa} used an implementation of the MEM that is restricted
to LO, and considered the $Z\gamma$ decay in an effective field
theory approach. The authors found only a marginal improvement between the 
MEM and a simpler approach that only used $m_{\ell\ell\gamma}$ as the discriminant. 
With the recent CMS study to guide us~\cite{CMS:Zgam}, the aim of this paper is to re-investigate the channel
using the MEM@NLO algorithm and the full loop matrix element for the
decay.

This paper proceeds as follows. In section~\ref{sec:HZga} we discuss the $H\rightarrow Z\gamma$ calculation and the form of the matrix elements
used in our analysis. Section~\ref{MEM@NLO} provides a brief overview of the MEM@NLO technique and we present our results in
section~\ref{sec:results}. Finally in section~\ref{sec:conc} we draw our conclusions.

\section{The Higgs decay to $Z\gamma$} 
\label{sec:HZga}
 
In this section we briefly discuss the calculation of $H\rightarrow Z\gamma$ as it is implemented in the code MCFM~\cite{Campbell:1999ah,Campbell:2011bn}. 
The $H\rightarrow Z\gamma$ decay was first considered over twenty years ago~\cite{Cahn:1978nz,Bergstrom:1985hp}.
We consider the process,
\begin{equation}
H(p_0) \rightarrow \ell^+(p_3)+\ell^-(p_4) + \gamma(p_5) \;,
\end{equation}
where the momenta are shown in parentheses.
The squared matrix element for this loop-induced process has the following form, 
\begin{eqnarray}
|M_{H\rightarrow\ell^+\ell^-\gamma}|^2 =\frac{e^8s_{34}(s_{35}^2+s_{45}^2)
}{2\sin^2{\theta_W}(16\pi^2m_W)^2}(|\mathcal{F}_L|^2+|\mathcal{F}_R|^2). \nonumber\\
\end{eqnarray}
In this expression we have introduced the electroweak coupling $e$ and the weak mixing angle ${\theta_W}$ and kinematic invariants 
are defined through $s_{ij}=(p_i+p_j)^2$. 
The left-handed and right-handed amplitudes are defined by,
\begin{eqnarray}
\mathcal{F}_{L,R} &=& \frac{4 Q_t N_c m_t^2}{s_{34}}\bigg(Q_tQ_{\ell}+\frac{1}{2}(v^t_L+v^t_R)v_{L,R}\mathcal{P}_Z\bigg)F_t\nonumber\\&&+\bigg(Q_{\ell}+v_{L,R}^{\ell}\cot{\theta_W}\mathcal{P}_Z\bigg)F_W,
\end{eqnarray}
in terms of the charge of the leptons and top quarks ($Q_\ell$, $Q_t$) the number of colours $(N_c)$ and the top mass $(m_t)$.
The vector couplings are defined as, 
\begin{eqnarray}
v_{L}^{\ell} &=& \frac{-1-2Q_{\ell}\sin^2{\theta_W}}{2\sin{2\theta_W}},  \; \; v_{R}^{\ell}=-\frac{2Q_{\ell}\sin^2{\theta_W}}{\sin {2\theta_W}},\\
v_{L}^{t}&=&  \frac{1-2Q_{t}\sin^2{\theta_W}}{2\sin{2\theta_W}}, \;\; v_{R}^{t}=-\frac{2Q_{t}\sin^2{\theta_W}}{\sin {2\theta_W}}.
\end{eqnarray}
Finally the function $\mathcal{P}_Z$ describes the $Z$ propagator (with width $\Gamma_Z$), 
\begin{eqnarray}
\mathcal{P}_Z=\frac{s_{34}}{s_{34}-m_Z^2+i\Gamma_Zm_Z}.
\end{eqnarray}
The loop integral functions are contained in $F_W$ and $F_t$, which are defined by, 
\begin{eqnarray}
F_W&=& 2 \Bigg[\frac{s_{345}}{m_W^2}\left(1-2\frac{m_W^2}{s_{34}}\right)\nonumber\\&&+2 \left(1-6\frac{m_W^2}{s_{34}}\right)\Bigg] C_2(p_5,p_{34},m_W,m_W,m_W) \nonumber \\
     &+& 4 \left(1-4\frac{m_W^2}{s_{34}}\right)  C_0(p_5,p_{34},m_W,m_W,m_W) \;,
\end{eqnarray}
and,
\begin{eqnarray} 
F_t&=&C_0(p_5,p_{34},m_t,m_t,m_t)\nonumber\\&&+4 C_2(p_5,p_{34},m_t,m_t,m_t) \;.
\end{eqnarray}
In these expressions
$C_0$ and $C_2$ are standard Passarino-Veltman tensor integrals. Further details can be found in Refs.~\cite{Gunion:1989we,Gunion:1992hs,Djouadi:1996yq}.

In addition to the decay described above, MCFM contains NLO Higgs and QCD $Z\gamma$ production, including the $gg\rightarrow Z\gamma$ loop induced 
process~\cite{Campbell:2011bn}. We will use these matrix elements to calculate our weights. 

\section{The MEM@NLO technique} 
\label{MEM@NLO}

This section provides a brief overview of the MEM technique developed
in Ref.~\cite{Campbell:2012cz}, to which we refer the interested reader
for a more complete discussion. The crux of
the MEM method is to provide an event--by--event weight using the matrix
element. At LO one defines each event to be weighted by
the following quantity
\begin{eqnarray}
\tilde{P}_{LO}(\tilde{\phi})=\frac{1}{\sigma_{LO}} \int dx_1 \, dx_2 \,d\phi  \,\delta(x_1x_2s-Q^2) \nonumber\\\times f^{j}(x_1)f^{i}(x_2) \mathcal{B}_{ij}(x_1,x_2,\phi) W(\phi,\tilde{\phi}).
\label{eq:LOMEM}
\end{eqnarray} 
Here $f(x)$ represent the PDF with momentum fraction $x$, $Q^2$ is the invariant mass of the EW final state, $\mathcal{B}_{ij}$ represents the LO matrix element, which depends on the final state phase space point $\phi$, which is derived from the input event from data ($\tilde{\phi}$), via the transfer function $W(\phi,\tilde{\phi})$. The weights are normalized by the LO cross section $\sigma_{LO}$. Often in this paper we will use the following weight, which is defined in the limit of a perfect detector setup, i.e $W(\phi,\tilde{\phi})=\delta(\phi-\tilde{\phi})$, 
\begin{eqnarray}
{P}_{LO}({\phi})=\frac{1}{\sigma_{LO}} \int dx_1 \, dx_2   \,\delta(x_1x_2s-Q^2) \nonumber\\\times f^{j}(x_1)f^{i}(x_2) \mathcal{B}_{ij}(x_1,x_2,\phi).
\label{eq:plo}
\end{eqnarray} 
This weight has the advantage of requiring fewer Monte Carlo integrations, and hence it is less computationally expensive.
However, one must be confident that the analysis is not sensitive to such a simplifying assumption. For instance, in this study the narrow 
width of the Higgs would spoil this assumption, since any event with $m_{\ell\ell\gamma} = m_H$ would result in a large weight compared to the 
remaining events in the sample, yielding unrealistic results. Therefore in order to use the above definition, care must be taken with variables (in this case the invariant mass) 
that are extremely sensitive to detector resolution. We shall discuss this further in the next section.

The observed EW final state typically recoils against hadronic activity that is not modeled in the leading order calculation. 
In order for the weights to be well-defined and unique
one must therefore perform a boost to ensure that the final state $\phi$ is balanced in $p_T$. Then one can apply the PDF weighting assuming two
beams colliding in the $z-$direction. Since there are multiple Lorentz transformations satisfying these requirements which are connected by longitudinal boosts
to each other, we integrate over the allowed range. This results in the corresponding integration over $x_1$ (or $x_2$) in eq.~(\ref{eq:LOMEM}). We refer to the
set of $p_T$ balanced final state frames collectively as the MEM frame. We note that failure to perform this boost and subsequent integration results in a either an ill-defined (no-boost) or
non-unique (no integration over boosts), and hence theoretically unreliable weight. 

The MEM frame allows calculation of weights accurate to NLO defined as~\cite{Campbell:2012cz},
\begin{eqnarray}
\tilde{P}_{NLO}(\tilde{\phi})=\frac{1}{\sigma_{NLO}}\int d\phi (V(\phi)+R(\phi))W(\phi,\tilde{\phi}).
\label{eq:pnlo}
\label{eq:NLOMEM}
\end{eqnarray} 
The virtual corrections are expressed as, 
\begin{widetext}
\begin{eqnarray}
V{(\phi})= \int dx_1 \, dx_2   \,\delta(x_1x_2s-Q^2) f^{j}(x_1)f^{i}(x_2) \hat{\mathcal{V}}_{ij}(x_1,x_2,\phi),
\end{eqnarray} 
\end{widetext}
where $\hat{\mathcal{V}}$ represents the contributions from the Born matrix element, the interference between the Born and one-loop amplitudes
and the integrated form of a relevant subtraction term (in this work we use a slightly modified Catani-Seymour~\cite{Catani:1996vz} dipole approach).
The real radiation pieces involve integration over an unresolved emission for which we use the forward-branching phase space (FBPS) generator described in ref~\cite{Giele:1993dj,Giele:2011tm}
\begin{widetext}
\begin{eqnarray}
R{(\phi})= \int dx_1 \, dx_2 \, d\phi_{FBPS} \,\delta(x_1x_2s-Q_{FBPS}^2) f^{j}(x_1)f^{i}(x_2) \hat{\mathcal{R}}_{ij}(x_1,x_2,\phi,\phi_{FBPS}).
\end{eqnarray} 
\end{widetext}
In the above $\hat{\mathcal{R}}_{ij}$ represents the matrix element for the
Born amplitude plus one additional parton, rendered finite by the
corresponding subtraction terms. Note that the constraining delta
function for the PDF's has changed definition with respect to LO, $Q^2_{FBPS}$ is the invariant mass of the Born 
final state plus the NLO emission. For
full details of the FBPS and subtraction setup we refer the reader to
ref.~\cite{Campbell:2012cz}.

The main difference between the LO and NLO MEM is the
integration over the real phase space. Some events in the lab frame,
when mapped to the MEM frame, no longer lie in the fiducial region
defined by the lab frame cuts (which the weights are normalized to)
and therefore are assigned zero weight~\cite{Campbell:2012cz}. At NLO
these events can have non-zero weights since the real emission
contributions can boost these events back into the fiducial region. A
simple way of interpreting this phenomenon is that NLO covers a larger
kinematic phase space than LO. This larger phase space manifests
itself as an ability to accept events which do not possess the correct
kinematics to have arisen from our LO discriminant. This is one of the
primary advantages of the MEM@NLO method compared to the MEM@LO (in
addition to the usual increase in confidence in the understanding of
the theoretical systematic error from using an NLO
prediction). Typically one finds that $\mathcal{O}(\alpha_S)$ (i.e. of
order 10\%) of the events fail the LO cuts.

Each event in the data set can now be assigned a unique LO or NLO
weight associated with a theoretical hypothesis controlled by the
underlying matrix element. In our case we will assign it a weight based
on the signal matrix element $P^{H\rightarrow Z\gamma}$ or the irreducible background
production of $Z\gamma$, $P^{Z\gamma}$. One can use these individual
quantities to build discriminants. Unless stated otherwise, our default is to use NLO matrix elements in our weight calculations.

\section{Results} 
\label{sec:results}
In order to study the MEM@NLO for the $H\rightarrow Z\gamma$ decay
mode we generate samples of signal and background events. We do this
using the SHERPA event generator~\cite{Gleisberg:2008ta}. For the
background we generate a CKKW~\cite{Catani:2001cc} matched sample of $Z\gamma$ events. For
the signal events we generate NLO matched Higgs events for $m_H=125 $ GeV. These Higgs
events are then subsequently decayed to the $Z\gamma$ final state
using the MCFM implementation that is described in Section~\ref{sec:HZga}. Since the Higgs is a scalar particle, production and
decay are uncorrelated. This allows us to simply calculate the decay
using MCFM in the rest frame of the Higgs and then boost it so that it 
has the four-momentum of the SHERPA event.
Throughout our studies we will use the CT10 PDF set~\cite{Lai:2010vv}.

The above procedure produces events at the particle level. However, in
order to study the light Higgs in a meaningful way one must include
some kind of detector simulation. This is because the light Higgs has
such a narrow width that the $m_{\ell\ell\gamma}$ spectrum is
dominated by the detector resolution. For example, the CMS technical design report~\cite{CMS:TDR}
estimates a resolution of photon energy using the following,
\begin{eqnarray}
\bigg(\frac{\Delta E_{\gamma}}{E_\gamma}\bigg) = \frac{3.6 \%}{\sqrt{E_\gamma/{\rm{GeV}}}} \oplus \frac{ 18.5\%}{E_\gamma/{\rm{GeV}}}\oplus 0.66\%
\label{eq:Esmear}
\end{eqnarray}
where $\Delta E_{\gamma}$ represents the width of the Gaussian smearing and $\oplus$ indicates that the quantities are to be added in quadrature. 
At $E = 30 $ GeV this provides a width of around 0.3 GeV. Using this smearing (and the equivalent leptonic quantity) we find a Higgs boson lineshape 
which is too narrow compared to that recently reported by CMS~\cite{CMS:Zgam}, where the effective Gaussian width quoted is around $3-4$~GeV. Therefore 
in order to match onto the results of this paper we inflate our Gaussian smearing to, 
\begin{eqnarray}
\Delta E_{\gamma} = 2 \; {\rm{ GeV}}, \quad \Delta E_{\ell} = 0.5 \; {\rm{ GeV}}.
\end{eqnarray}
Our resulting lineshape for the Higgs is now in good agreement with ref.~\cite{CMS:Zgam}. Note that, since our enhanced width is around a factor of six greater 
than that arising from the energy dependent piece eq.~(\ref{eq:Esmear}), we drop the energy dependence for simplicity. 

After smearing our events as described above we apply the following lab frame cuts, 
\begin{widetext}
\begin{eqnarray}
p_T^{\gamma} > 15\; {\rm{GeV}}, \; \;  |\eta^{\gamma}| < 2, \; \;  p_T^{\ell} > 20\; {\rm{GeV}}, \; \;  |\eta^{\ell}| < 2,   \; \;
60 < m_{\ell\ell} < 120~{\rm{GeV}},   \; \; 115 < m_{\ell\ell\gamma} < 135~{\rm{GeV}},  \; \; R_{\ell\gamma} > 0.7 \;.
\end{eqnarray}
\end{widetext}

Note that we have kept the cuts on the lepton and photon $p_T$ loose.  Part of
the attraction of the MEM discriminant is that it will naturally
select events which have the correct kinematics to be signal events,
therefore one does not have to spend time attempting to optimize
kinematic selection criteria. Of course, if some observable clearly
discriminates signal from background, cuts on this quantity should be
applied in order to reduce the overall computational load. For this
reason we impose a tight cut on the invariant mass of the $Z\gamma$
system, centered on $125$~GeV.

\subsection{Generation of reducible background events.} 

In order to effectively simulate LHC conditions we must also consider
events which do not arise from the irreducible $Z\gamma$ background,
but instead  are misidentified in the detector. Since they are naturally very dependent on
the exact detector setup and modeling, such events are difficult to
accurately simulate in our study. However since they are a large
fraction of the resulting event sample~\cite{CMS:Zgam}, it is
necessary to attempt to provide an estimate of our discriminant on a
``fake" sample. To this end we generate a sample of fake events in the
following way. We assume that the dominant component of the fake
events results from $Z+$jet events in which the leptons from the $Z$
decay are clean, but the jet is mismeasured as a photon. As a crude model, we use
SHERPA to generate $Z+$jet events and then smear the $p_T$ and $\eta$ of
the (leading) jet by Gaussian functions with a width of $10$~GeV and $0.5$ respectively.
Our event sample is then generated by applying the cuts described above in the previous section to the smeared events, treating the 
smeared jet as a photon.

\subsection{The MEM@NLO as a kinematic discriminant} 

We first discuss our definition of the Higgs signal hypothesis, which is particularly
important because of the very narrow SM width. One approach is to define a weight for each event
by integrating over transfer functions that model the detector resolution, as in eq.~(\ref{eq:LOMEM}).
This approach allows one to test a single Higgs mass hypothesis for a given set of events,
but requires additional integrations per event. An
alternative approach is to change the Higgs mass hypothesis on an event-by-event basis
by choosing $m_H=m_{\ell\ell\gamma}$. In this
scenario one effectively changes the propagator in the matrix element to,
\begin{eqnarray} 
\frac{1}{(s-m_H)^2+i\Gamma_H m_H} \rightarrow 
\frac{1}{i\Gamma_H m_{\ell\ell\gamma}} \;.
\end{eqnarray} 
Since $\Gamma_H$ is very small this approach makes each event have a
large $P_S$ and the collection of events can no longer define a probability density function.
In addition $P_S \gsim P_B$ even for background event samples. However, one
still expects discriminating power between signal and background since $P_S$ arising from the 
events which match onto the signal hypothesis will be larger than that for $P_S$ from the background. 
Finally we note that the (signal) normalization is defined uniquely for each event by $\sigma(m_H=m_{\ell\ell\gamma})$.  

This technique has been used extensively in studies
involving kinematic discriminants in $H\rightarrow ZZ\rightarrow
4\ell$ \cite{CMS:2012br,Avery:2012um}. The advantage of this
technique is that there are less integrations per event and thus the
weights are computationally less expensive. One can then restore the
invariant mass $m_{\ell\ell\gamma}$ as an additional discriminant since Higgs events will cluster in invariant mass
whilst the background will be more diffuse. We will adopt this approach for the remainder of this paper.

With the event samples generated as described in the
previous sections we can now introduce our
discriminant $\mathcal{D}$. 
There is a range of possibilities but in
this paper we will choose,
\begin{eqnarray}
\mathcal{D}=-\log{\left(\frac{P_{B}}{P_S+P_B}\right)}.
\label{eq:Ddefn}
\end{eqnarray}
Events which arise from background should have larger $P_B$ than $P_S$ so the ratio in the logarithm is near one. As a result,
events with $\mathcal{D}$ nearer zero should be more background-like than signal.

\begin{figure*}
\begin{center}
\includegraphics[width=14cm]{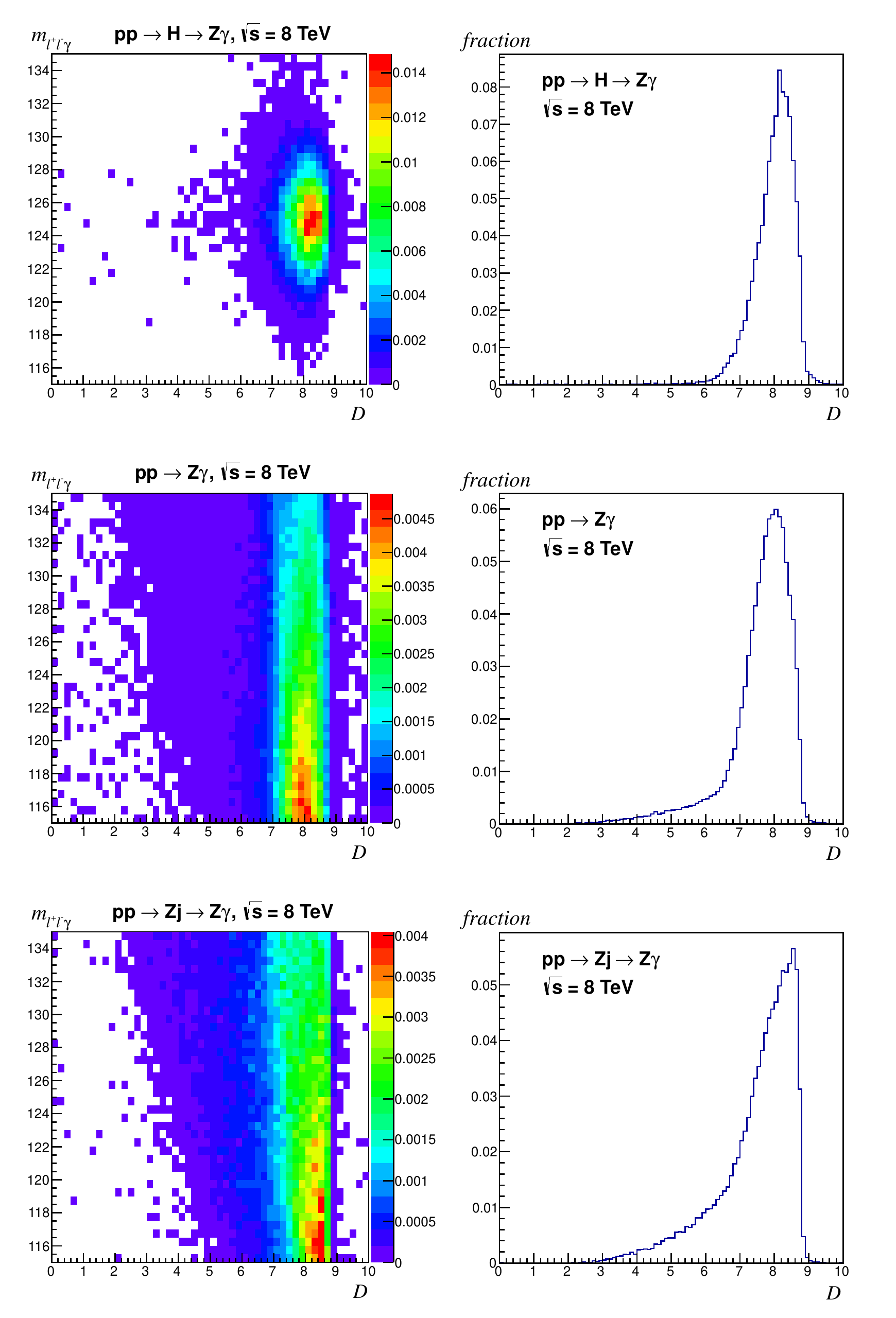}
\caption{Distribution of events in terms of the invariant mass of the final state, $m_{\ell\ell\gamma}$ and our discriminant $\mathcal{D}$ defined in eq.~(\ref{eq:Ddefn}).
The two-dimensional histograms (left) present the density of events in the plane of $\mathcal{D}$
and $m_{\ell\ell\gamma}$. The right hand panels represent the distribution of $\mathcal{D}$ for our
signal (top row), background (middle row) and fake (bottom row) samples.  Each 
sample is normalized to the total number of events in the sample.} 
\label{fig:disc_all}
\end{center}
\end{figure*}
We present results for $\mathcal{D}$ for our three different event
sample classes in Fig.~\ref{fig:disc_all}.
The results are shown as two-dimensional histograms, binned both by the discriminant $\mathcal{D}$
and the invariant mass, $m_{\ell\ell\gamma}$. In addition, we also show one-dimensional projections of
these histograms, as a function of  $\mathcal{D}$ only.
As expected the signal
events peak at larger $\mathcal{D}$ than the corresponding background
distribution. The background and fake samples have roughly similar shapes (indicating some of the 
similarities between $Z$+jet and $Z\gamma$). Although the signal shapes are similar to the background, there are still significant
regions that are only populated by background events (but which may still have an invariant mass in the Higgs window).
In particular, both the background and fake samples have significant tails in the lower $\mathcal{D}$ region, whereas 
the signal sample does not.  For example, there are barely any $(~\sim 0.5\%)$ of signal
events with $\mathcal{D} < 6$. On the other hand around $10\%$ of the background events
lie in this region. Cutting at
$\mathcal{D} > 6$ would thus be an almost zero-cost reduction in signal at the
expense of a non-trivial background number of events. 

The scales on the two-dimensional histograms illustrate the stark differences between the
signal and background events in the ($\mathcal{D}, m_{\ell\ell\gamma}$) plane. The area of highest density for 
the signal events (around the truth value, $m_{\ell\ell\gamma} = 125$~GeV) is around three times
greater than the corresponding highest density region for the background (which is at much lower invariant mass).
Retaining only the events that satisfy $\mathcal{D} > 7$, one rejects $21\%$ of 
the irreducible background events and keeps $93\%$ of the signal. A higher cut, $\mathcal{D} > 8$ rejects
$64\%$ of the background and retains $55\%$ of the signal.
In an experimental analysis one would thus choose the optimal value of $\mathcal{D}$ at which to cut in order to optimize 
the signal to background ratio.  Since our model of the fakes is less developed than our signal and
background models we optimize our cut on the discriminant on the
combination of signal and irreducible background samples. We find  
a value of the cut at $\mathcal{D} > 7.5$ corresponds to a
signal efficiency of  81\%, with an background rejection of 37\%.
We note that here we have chosen a fairly simple cut on $\mathcal{D}$ that optimizes
$S/\sqrt{B}$. One could instead
perform counting experiments using contours in the ($\mathcal{D}, m_{\ell\ell\gamma})$ plane although
such a study is beyond the scope of this work.

\begin{figure*}
\begin{center}
\includegraphics[width=14cm]{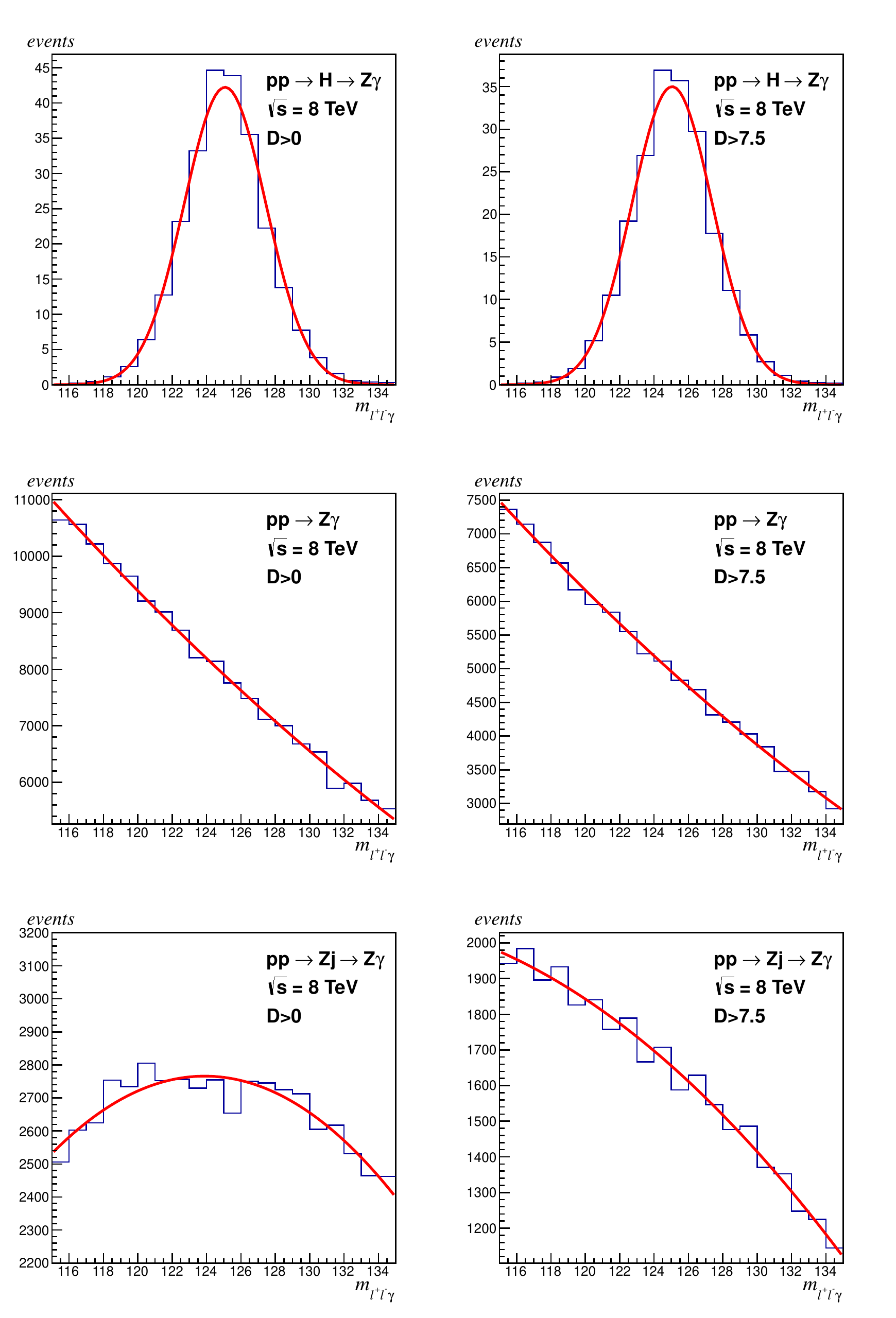}
\caption{Invariant mass distributions (for $m_{\ell\ell\gamma}$) before (left panels) and after (right panels) our analysis cut $\mathcal{D}>7.5$. 
Distributions are shown for the signal events (top row), the irreducible background (middle row) and fakes (final row). The number of events in each
distribution is normalized to the irreducible background sample without application of any cuts on $\mathcal{D}$, as described in the text.
The red curve indicates a Gaussian (polynomial) fit to the signal (background) data.} 
\label{fig:DSvDB_cut}
\end{center}
\end{figure*}
We plot the invariant mass $m_{\ell\ell\gamma}$ before and after our cut ($\mathcal{D} > 7.5$) 
for our three samples in Fig.~\ref{fig:DSvDB_cut}. Each sample is weighted to reflect the number
of events expected, given the total number of irreducible background $Z\gamma$ events.  We weight
our signal sample by the ratio of cross sections (including a NLO to
NNLO $K$-factor of 1.2~\cite{Anastasiou:2012kq}).
We normalise our fake sample to be compatible with that reported by CMS~\cite{CMS:Zgam}, namely by fixing the number of fake events to be one third of the irreducible background.  
Our cuts have altered the shape of the background and fake samples, whilst maintaining
the overall shape of the signal.

Ultimately we would like to investigate the efficiency of this method in the vicinity of
the Higgs signal at $m_H=125$~GeV. We therefore define a window,
\begin{equation}
122 < m_{\ell\ell\gamma} < 128~{\rm GeV} \;,
\end{equation}
where the width has been optimized for the analysis below.
We attempt to quantify the
improvement the cut on $\mathcal{D}$ has made in the following way.
We define the quantity,
\begin{eqnarray}
\alpha = \frac{\sqrt{N_{Z\gamma}+N_{fakes}}}{N_H}\;,
\end{eqnarray}
where $N_{X}$ represents the expected number of events for process $X$.
Our measure includes no treatment of systematic errors and instead only assumes the $S/\sqrt{B}$
scaling of the statistical uncertainty. In spite of its shortcomings
compared to the full experimental analysis, $\alpha$ can provide us
with an estimate of the improvement one might envisage after applying
our cut. We find,
\begin{eqnarray}
\frac{\alpha_{D> 0}}{\alpha_{D > 7.5}} = 1.52.
\end{eqnarray}
Since $\alpha$ scales as $\mathcal{L}^{-1/2}$, using a cut of $D > 7.5$ is (statistically) equivalent to taking 2.31 times more data.  

Before concluding, we will briefly consider the impact of using the leading order method,
MEM@LO rather than the NLO one.  We find that the fraction
of events which fail the fiducial cuts at LO is larger for the Higgs signal than for
the irreducible background. As a result the MEM@NLO produces slightly better
signal over background ratios than the MEM@LO.
For example we find,
\begin{eqnarray}
\frac{\alpha^{LO}_{D> 0}}{\alpha^{LO}_{D > 7.5}} = 1.41,
\end{eqnarray}
which is 7\% smaller than the corresponding NLO value.
This small difference, between the LO and NLO analyses, indicates that the method is perturbatively stable and the
theoretical systematic uncertainty is under good control.

\section{Conclusions} 
\label{sec:conc}

In this paper we have presented an application of the MEM@NLO to
searches for the Higgs boson in the decay channel $Z\gamma$. This
channel is extremely challenging experimentally as can be seen from
the preliminary results from CMS~\cite{CMS:Zgam}. The reasons for
these difficulties are two-fold. Firstly the $H\rightarrow Z\gamma$
branching ratio is already very small, even before the requirement that 
the $Z$-boson decays to muons and electrons only. Since the background production
of $Z$ in association with a photon is large, one naturally has low
signal to background ratios. Secondly the kinematics of the decay for
a Higgs boson with mass $m_H\sim 125$~GeV force the final state photon to have a
relatively soft $p_T$. The matrix element has a soft singularity as
$p_T^{\gamma} \rightarrow 0$ and therefore the background is very large
in the region in which the Higgs signal peaks.
Once detector effects are included there is very little
difference between the signal and background events in the transverse
variables.

Given these difficulties, it is essential to utilize all the remaining
differences between the signal and background processes. One approach,
the matrix element method, uses a theoretically defined matrix element to assign a weight
to each experimental event. When there is a good match
between the theoretical hypothesis and the input events the weights
become larger. Therefore one can use the MEM to produce samples of
events which increase the signal to background ratio for a certain theoretical hypothesis. 
Recently the MEM
has been extended to NLO for electroweak final
states~\cite{Campbell:2012cz}. We used this MEM@NLO to calculate
signal and background discriminants for a sample of events generated using SHERPA.
Our event sample included showered and hadronized Higgs
signal and SM background events as well as a crude model of $Z+$jet fake
events. Higgs decays to $Z\gamma$ were generated using the MCFM
implementation. Our estimates of resolution effects and fake rates were guided by the
recent results from CMS presented in Ref.~\cite{CMS:Zgam}.

We used the MEM@NLO to construct a discriminant $(\mathcal{D})$ from the event-by-event weights 
$P_S$ (using the signal matrix element) and $P_B$ (the background matrix element). In defining these 
weights we removed the invariant mass as a discriminating variable. As a result we were 
subsequently able to create a two-dimensional discriminant in $\mathcal{D}$ and $m_{\ell\ell\gamma}$. 
In this plane the signal events cluster around $m_H$ and at higher $\mathcal{D}$ compared to those 
arising from the background and fakes. Therefore, by cutting on $\mathcal{D}$ and $m_{\ell\ell\gamma}$,
we were able to improve our measure of the signal significance, $S/\sqrt{B}$.  We found 
that $S/\sqrt{B}$ increased by around a factor of $1.5$ compared to the value obtained without any cut on $\mathcal{D}$,
suggesting that roughly half as much data
would be needed to obtain the same limit on $H\rightarrow Z\gamma$. 
We found that the MEM@LO algorithm is also able to provide $S/\sqrt{B}$ improvements by a factor of around $1.4$,
approximately 10\% less efficient than the MEM@NLO.

This search has provided an example of the power of the matrix element method
in a worst case scenario for a traditional analysis. We hope that the ideas presented in this paper are useful
to our experimental colleagues in the hunt for the Higgs boson in this
difficult channel. Code which calculates the weights described in
this paper is available upon request.

\section*{Acknowledgements} 

Fermilab is operated by Fermi Research Alliance, LLC under
Contract No. DE-AC02-07CH11359 with the United States Department of Energy. 

\bibliography{HZGam}

\end{document}